\documentstyle[aps,12pt,epsf]{revtex}
\def\beq{\begin{equation}}
\def\eeq{\end{equation}}
\def\bea{\begin{eqnarray}}
\def\eea{\end{eqnarray}}

\def\vel{\left|}
\def\ver{\right|}

\def\ga{\left(}
\def\dr{\right)}

\def\ba{\begin{array}}
\def\ea{\end{array}}

\def\bos{\lower 0.5cm\hbox{{\vrule width 0pt height 1.3cm}}}
\def\aaa{\lower 0.cm\hbox{{\vrule width 0pt height .8cm}}}
\def\dol{\lower 0.6cm\hbox{{\vrule width 0pt height .8cm}}}

\newcommand\pubdate{\today}

\newcommand\hepnumber{hep-ph/01xxx}
\def\csumb{
$^a$ Institute of Theoretical Physics, Academia Sinica, \\
      P.O.Box 2735, Beijing 100080, P.R.China \\
$^b$ Institut f\"ur Theoretische Physik,
Universit\"at Karlsruhe,   \\ D-76128 Karlsruhe, Germany}

\def\Title#1{\begin{center} {\Large\bf #1 } \end{center}}
\def\Author#1{\begin{center}{ \sc #1} \end{center}}
\def\Address#1{\begin{center}{ \it #1} \end{center}}

\newcommand\pubblock{\rightline{\begin{tabular}{l}
         \pubdate\\ \hepnumber \end{tabular}}}

\begin{document}
\begin{titlepage}
\pubblock

\Title{
Heavy quark polarizations of $e^+e^-\rightarrow q \bar q h$
in the general two Higgs doublet model
}

\Author{ Chao-Shang Huang $^a$, Shou-hua Zhu$^b$ }
\Address{\csumb}
\vfill

\begin{abstract}
The polarizations of the heavy quark ($q=t$ or $b$)
in the process $e^+e^- \rightarrow q \bar q h$ have been calculated
in the general two Higgs doublet model. The CP violating normal polarization
of the top quark can reach $8 \%$, and $2 \sim 3\%$ for the
bottom quark, while it is zero in the standard model. The
longitudinal and transverse polarizations of the bottom quark  can be significantly
different from those in SM and consequently  could
aslo be used as the probe of the new physics. 
\end{abstract}

\vfill
\end{titlepage}
\eject \baselineskip=0.3in

Searching for a Higgs boson has been a major motivation
for many current and future collider experiments, since
Higgs bosons encode the underlying physics of mass generation.
The Next Linear Collider (NLC) running at center-of-mass (CMS)
 energies from 0.5
to 2 Tev is designed to examine the nature of the Higgs bosons
in detail. The present lower bounds on Higgs boson
mass from the direct searchs are $m_h>114.1$ Gev (at 95$\%$ CL)
\cite{lb1} for the standard model (SM) Higgs boson, and $m_h>91.0$
Gev and $m_A>91.9$ Gev (at 95$\%$ CL)~\cite{lb2} for the light
neutral scalar and pseudoscalar Higgs bosons of the minimal 
supersymmetric standard model (MSSM). In the MSSM the upper bound
of the light neutral scalar Higgs boson mass is predicted to be
about 130 Gev~\cite{lb3}. Therefore, it is possible and expected
to examine the properties of the Higgs bosons in 
the process $e^+e^-\rightarrow q \bar{q} h$ (q=t, b) at NLC.

It has been estimated that a detailed investigation of cross-sections
of a few fb is possible at NLC with a yearly integrated luminosity of 
the order of 100 $fb^{-1}$~\cite{cs}. The associated production of a
Higgs boson with a pair of heavy fermion-antifermion is of a good place to
study the couplings of Higgs bosons to fermions, in particular, to
probe new Higgs-fermion interactions at the tree level. The possibility
of observing large signitures of new CP-violating and flavor-changing
Higgs-quark couplings in NLC experiments
has been discussed~\cite{sbs}. 

If there are CP violating couplings in the extended Higgs sector (e.g., in a general two Higgs
model, MSSM etc.) CP violation in the process $e^+e^-\rightarrow q \bar{q} h$ arises because 
of interference of the diagrams where the Higgs 
bosons is coupled to a Z boson with the diagrams where the Higgs boson is radiated off the 
quark or anti-quark.
  There are two kinds of CP-odd observables that can trace the tree level
CP effects in the process $e^+e^-\rightarrow q \bar{q} h$. One is
consisted of the triple momenta
${\bf O}=\vec{p}_{-}\cdot(\vec{p}_q\times \vec{p}_{\bar{q}})$, the other
is of the mixed triplet of momenta and spin $P_N=\vec{s}_q
\cdot(\vec{p}_{-}\times \vec{p}_q)$, i.e., the normal polarization of the 
quark. The formmer is examined in $e^+e^-\rightarrow t\bar{t} h$ in 
ref.~\cite{sbs}, and we shall study the latter in the paper. It is known that
in the semi-leptonic decay $B \rightarrow X_s l^+ l^-$, $P_N$ is the 
CP-violating projection of the lepton spin onto the normal of the decay plane. 
Because $P_N$ in $B \rightarrow X_s l^+ l^-$ comes from both the quark and 
lepton sectors, purely hadronic and leptonic CP-violating observables,
such as $d_n$ or $d_e$, do not necessarily strongly constrain $P_N$
\cite{add1}. So it is advantageous to use $P_N$ to investigate
CP violation effects in some extensions of SM \cite{add3}. For $e^+e^- 
\rightarrow q\bar{q} h$,  $P_N$ can be defined as the 
CP-violating projection of the quark spin onto the normal of the plane 
determined by $\vec{p}_{e^-}$ and $\vec{p}_q$. 
Although there is no advantage mentioned above, 
it is useful to investigate CP effects by measuring $P_N$
since it can give more information on the couplings of Higgs to fermions.
We find that the normal polarization $P_N$ of the heavy quark in $e^+e^-\rightarrow q \bar{q} h$
 is sensitive to CP violation effects due to the extended Higgs sector. Therefore, the two
kinds of CP-odd observables are complementary each other in probing the CP-odd couplings of
Higgs bosons to quarks. In the paper we shall investigate the polarizations of q (q=t, b) and
the cross section for the $q\bar{q}h$ production at NLC in a general two-Higgs-doublet model.


In the general two-Higgs-doublet model used here \cite{froggat},
the $h q \bar q$ interaction lagrangian piece ($h$ represents one
of the three neutral Higgs boson) can be written as:
\begin{equation}
{\cal L}= -\frac{g_W}{ 2 m_W} \left(m_t h \bar t ( a_t + i b_t
\gamma_5 ) t +  m_b h \bar b ( a_b + i b_b \gamma_5 )b \right)
\label{tthcoupling}~.
\end{equation}
If we choose $h=H_1$ as Ref. \cite{froggat}, $a$ and $b$ can be
expressed as
\begin{eqnarray}
a_{t};a_{b} &=&\frac{\cos\alpha_1}{\sin\beta}; ~
-\frac{\sin\alpha_1 \cos\alpha_2}{\cos\beta},\nonumber \\
b_t;b_{b} &=&\cot\beta \sin\alpha_1 \sin\alpha_2; ~ \tan\beta
\sin\alpha_1 \sin\alpha_2
 \label{mod2coup} ~
\end{eqnarray}
with $\tan\beta \equiv v_u / v_d$ and $v_u$($v_d$) is the
vacuum-expectation-value (VEV) responsible for giving mass to  the
up(down) quark. Three Euler angles $\alpha_{1,2,3}$
\cite{froggat} parameterize the neutral Higgs mixing matrix.
Note that in the SM, the only couplings 
 of the one neutral Higgs present, and $a_{t,b}^h=1, b_{t,b}^h=0$.

For the process $e^+(p_2) e^-(p_1) \rightarrow  h(k_1) t(k_2) \bar
t(k_3)$ \footnote{For the simplicity of presentation,
we give the formulas for $t \bar t h$ production. The results
of $b \bar b h$ production could be easily obtained by substituting
the couplings and kinematic variables.}, 
we define $(p_1+p_2)^2=S$. And the amplitude is as
following
\begin{eqnarray}
M&=&\frac{e^3}{4 c_w^4 s_w^3 m_w} \{ \overline{u}(k_2)\gamma_\mu
(f_1 P_R+ f_2 P_L ) v(k_3) \overline{v}(p2) \gamma_\mu P_R u(p_1)+
\overline{u}(k_2)\gamma_\mu( f_3P_R \nonumber
\\ &&+ f_4 P_L
) v(k_3) \overline{v}(p2) \gamma_\mu P_L u(p_1)   +  c_w^2 m_t [
\overline{u}(k_2)( f_5 P_R +f_6 P_L ) v(k_3) \overline{v}(p2)
\not{k}_2 P_R u(p_1)
 +
\overline{u}(k_2)( f_7 P_R  \nonumber
\\ &&+f_8 P_L ) v(k_3)
\overline{v}(p2) \not{k}_2 P_L u(p_1)
 +
  \overline{u}(k_2)( f_9 P_R +f_{10} P_L ) v(k_3)
\overline{v}(p2) \not{k}_3 P_R u(p_1)
 +
\overline{u}(k_2)( f_{11} P_R \nonumber
\\ && +f_{12} P_L ) v(k_3)
\overline{v}(p2) \not{k}_3 P_L u(p_1)
  + \frac{1}{2}
\left( \overline{u}(k_2)(\not{p}_1+\not{p}_2)\gamma_\mu P_R v(k_3)
\overline{v}(p2) \gamma_\mu (f_{13}P_R+ f_{14}P_L) u(p_1) \right.
\nonumber \\ &&\left.  +
\overline{u}(k_2)(\not{p}_1+\not{p}_2)\gamma_\mu P_L v(k_3)
\overline{v}(p2) \gamma_\mu (f_{15}P_R+ f_{16}P_L) u(p_1) \right)
 ] \},
\end{eqnarray}
where
\begin{eqnarray}
f_1 &=& \left( -{\it g_A^e} + {\it g_V^e} \right) \,
  \left( - F_2 A\,{{{\it c_w}}^2}\,{\it g_A^t}\,{{{\it m_t}}^2}  -
    F_4 A^*\,{{{\it c_w}}^2}\,{\it g_A^t}\,{{{\it m_t}}^2} +
    F_5\,{\it ( g_V^t- g_A^t)}\,{\it c}\,{{{\it m_w}}^2} \right),
\end{eqnarray}
$f_2$ is obtained by substituting $-g_A^t$ for $g_A^t$ and $A^*$ for A in $f_1$,
$f_3$ by substituting $-g_A^e$ for $g_A^e$ in $f_1$, and $f_4$ by substituting $-g_A^t$ for $g_A^t$, 
$A^*$ for A and $-g_A^e$ for $g_A^e$ in $f_1$,
\begin{eqnarray}
f_5&=&  F_2 A (g_A^t+g_V^t) (g_V^e-g_A^e)+
 F_5 {\it c}\,{\it \eta} (g_V^e-g_A^e) +
        4\,F_1 A\,{{{\it c_w}}^2}\,{\it e_t}\,{{{\it s_w}}^2}
\end{eqnarray}
$f_6$ is obtained by substituting $-g_A^t$ for $g_A^t$ and $A^*$ for A in $f_5$,
$f_7$ by substituting $-g_A^e$ for $g_A^e$ in $f_5$, and $f_8$ by substituting $-g_A^t$ for $g_A^t$, 
$A^*$ for A and $-g_A^e$ for $g_A^e$ in $f_5$,
\begin{eqnarray}
f_9&=&  F_4 A (g_A^t-g_V^t) (g_V^e-g_A^e)+
 F_5 {\it c}\,{\it \eta} (g_V^e-g_A^e) -
        4\,F_3 A\,{{{\it c_w}}^2}\,{\it e_t}\,{{{\it s_w}}^2}
\end{eqnarray}
$f_{10}$ is obtained by substituting $-g_A^t$ for $g_A^t$ and $A^*$ for A in $f_9$,
$f_{11}$ by substituting $-g_A^e$ for $g_A^e$ in $f_9$, and $f_{12}$ by substituting $-g_A^t$ for $g_A^t$, 
$A^*$ for A and $-g_A^e$ for $g_A^e$ in $f_9$,
\begin{eqnarray}
 f_{13}&=&
 A
 \left( (g_V^e- g_A^e)[  F_2 (g_V^t+ g_A^t)+ F_4 ( g_V^t- g_A^t)]
  +
       4\,(F_1+F_3) \,{{{\it c_w}}^2}\,{\it e_t}\,{{{\it s_w}}^2}
       \right)
\end{eqnarray}
$f_{14}$ is obtained by substituting $-g_A^e$ for $g_A^e$ in $f_{13}$,
$f_{15}$ by substituting $-g_A^t$ for $g_A^t$ and $A^*$ for A in $f_{13}$,
$f_{16}$ by substituting $-g_A^t$ for $g_A^t$, 
$A^*$ for A and $-g_A^e$ for $g_A^e$ in $f_{13}$,
\begin{eqnarray}
A&=& a+ i b,
\end{eqnarray}
\begin{eqnarray}
\frac{1}{F_1}&=& S \left(S-2 k_2 . (p_1+p_2)\right),
\end{eqnarray}
\begin{eqnarray}
\frac{1}{F_2}&=& (S- m_z^2) \left(S-2 k_2 . (p_1+p_2)\right),
\end{eqnarray}
$1/F_3$ and $1/F_4$ are obtained by substituting $k_3$ for $k_2$ in $1/F_1$ and $1/F_2$
respectively,
\begin{eqnarray}
\frac{1}{F_5}&=& (S- m_z^2) (2 k_2 . k_3 +2 m_t^2-m_z^2).
\end{eqnarray}
In above equations, $c$ is the coupling constant of $z^0-z^0-h$
($c=1$ corresponding to the SM case \cite{froggat})
\begin{eqnarray}
c &=& \sin\beta \cos\alpha_1 +\cos\beta \sin\alpha_1 \cos\alpha_3,
\end{eqnarray}
$g_A^e=1/2$, $g_V^e=1/2-2 s_w^2$, $g_A^t=1/2$, $g_V^t=1/2-4/3
s_w^2$, $e_t$ is the top quark charge, for top quark $\eta=1$ and
for bottom $\eta=-1$.

The next step is to square the amplitude while keeping top quark
spin information. Technically, we use $u \bar u =\frac{1}{2}
(\rlap/{p}+m)(1+\lambda \rlap/{s} \gamma_5)$, where $s$ is the
lorentz spin vector and $\lambda=\pm 1$. Because expressions are
lengthy, we don't show it here explicitly.
Let us now discuss the quark polarization
effects. We define three orthogonal unit vectors:
\begin{eqnarray} \vec{e}_L &=& \frac{\vec{p}_t}{\vel \vec{p}_t \ver}~,
 \nonumber
\\ \vec{e}_N&=& \frac{\vec{p}_{e^-} \times \vec{p}_t} {\vel
\vec{p}_{e^-} \times \vec{p}_t \ver}~, \nonumber \\ \vec{e}_T &=&
\vec{e}_N \times \vec{e}_L~, \nonumber \end{eqnarray} where
$\vec{p}_{e^-}$ and $\vec{p}_{t}$ are the three momenta of the
$e^-$  and the $t$ quark, respectively, in the center of mass of
the $e^+~e^-$ system. $P_L,~P_T$, and $P_N$, which correspond to
the longitudinal, transverse and normal projections of the top
spin, respectively, are defined as
\begin{eqnarray}
P_i (s) = \frac{ {\displaystyle{\sigma \ga \vec{n}=\vec{e}_i \dr -
\sigma \ga \vec{n}=-\vec{e}_i \dr}} } { {\displaystyle{\sigma\ga
\vec{n}=\vec{e}_i \dr + \sigma\ga \vec{n}=-\vec{e}_i \dr}} } ~.
\end{eqnarray}


In our numerical examples, we choose the $h$ as the lightest
mass Higgs and ignore other neutral Higgs contributions. 
Because we are mainly interested in the CP violation effects
for the process, as
a numerical example,
the mixing angles are taken as $\alpha_1=\alpha_2=\pi/4$ and
$\alpha_3=0$.

In Fig. 1-3, the top quark polarizations and cross sections
are presented as function of the Higgs boson mass, CMS energy and
$\tan\beta$. The cross sections for $t\bar{t}h$ have been given~\cite{sbs} and our
results are in agreemant with the previous ones. We see from the figures that the $P_N$ can 
reach $\sim 8\%$ and decreases with the increase of $\tan\beta$,
while $P_N$ is equal to zero in the SM. On the contrary, 
the longitudinal and transverse polarizations in the 2HDM 
don't change much compared to thoese in the SM, if the
Higgs mass in the 2HDM is equal to that in SM.

In Fig. 4-6, the bottom quark polarizations and cross sections
are presented as function of the Higgs boson mass, CMS energy and
$\tan\beta$. We can see that the $P_N$ increases with the
increment of $\tan\beta$ and can reach $2 \sim 3 \%$.
The longitudinal and transverse polarizations in the 2HDM 
can differ a lot compared to thoese in the SM. 

We would like to discuss the statistic significance in measuring the CP violating
observable $P_N$ with emphasis on estimating the sensitivity to $\tan\beta$. 
The statistic significance is defined by~\cite{sbs}
$N_{SD}=P_N \sqrt{L}\sqrt{\sigma}$, where $L$ is the effective luminosity for fully reconstructed
$q\bar{q}h$ events and $\sigma$ is the production cross section
for $e^+e^-\rightarrow q\bar{q} h$. We take $L=\epsilon {\cal L}$, where ${\cal L}$
is the total yearly integrated luminosity and $\epsilon$ is the overall efficiensy for
reconstruction of the $q\bar{q}h$ final state. For illustrative purposes, we choose 
${\cal L}=200 fb^{-1}$, $\epsilon=0.5$ (if $\epsilon=0.25$ our results would correspondingly
require 2 years of running), $m_h=150$ GeV and other model parameters same as those in
Fig. 3. For q=t, $P_N$ can be observed at a $1 \sigma$ level (which corresponds to 
$N_{SD}/\sqrt{L}=0.1$) for $\tan\beta \sim 0.6$ and at a 6.5 $\sigma$ level for $\tan\beta=0.1$.
For q=b, taking the values of the parameters as in Fig. 6,
even for $\tan\beta=50$, ${ L}=2000 fb^{-1}$ is needed in order
to observe $P_N$ at a 1 $\sigma$ level.

To summarize, the polarizations of the heavy quark ($t$ and $b$)
of the process $e^+e^- \rightarrow q \bar q h$ have been calculated
in the general two Higgs doublet model. 
We should note here that even for MSSM, the sizable Higgs sector
CP violation can be induced at one-loop, which can  manifest
themselves in the couplings of the lightest neutral Higgs boson
to the SM fermions \cite{Babu}.
Our numerical results show that
the normal polarization
of the top quark can reach $8 \%$, and $2 \sim 3\%$ for
bottom quark, which are zero in the SM. At the same time, the
longitudinal and transverse polarizations of the bottom quark can be significantly
different from those in SM and consequently could
aslo be used as the probe of the new physics for the process $e^+e^- \rightarrow
b  \bar b h$. As it is well-known, the top quark decays as a free quark since 
its lifetime is so short that it has no time to bind with light quarks before it
decays. Therefore, the top quark polarizations can be measured by measuring the energy
spectra of its decay W's, or of leptons from W decays~\cite{sp}. Furthermore,
a linear collider with $200 fb^{-1}$ will begin to be sensitive to 
$\tan\beta < 0.6 $,
for the $t\bar{t}h$ final state if the Higgs boson h is light (say, smaller than
or equal to 150 GeV) and the mixing angles of neutral Higgs bosons are not close to
the end points 0 and $\pi/2$. For b quark, its polarizations
can be measured by measuring those of its fragmentation $\Lambda_b$. Nevertheless, it is not
easy to observe due to the small probability (the probablity that a bottom quark fragmentizes into 
a $\Lambda_b$ is about 10 $\%$). 
\\

The author would like to thank G. Eilam and  S.~Bar-Shalom
for useful discussions.
This work was supported in part by the Alexander von Humboldt
Foundation and
National Nature Science Foundation of China.

\newpage

\begin{figure}
\epsfxsize=12 cm
\centerline{\epsffile{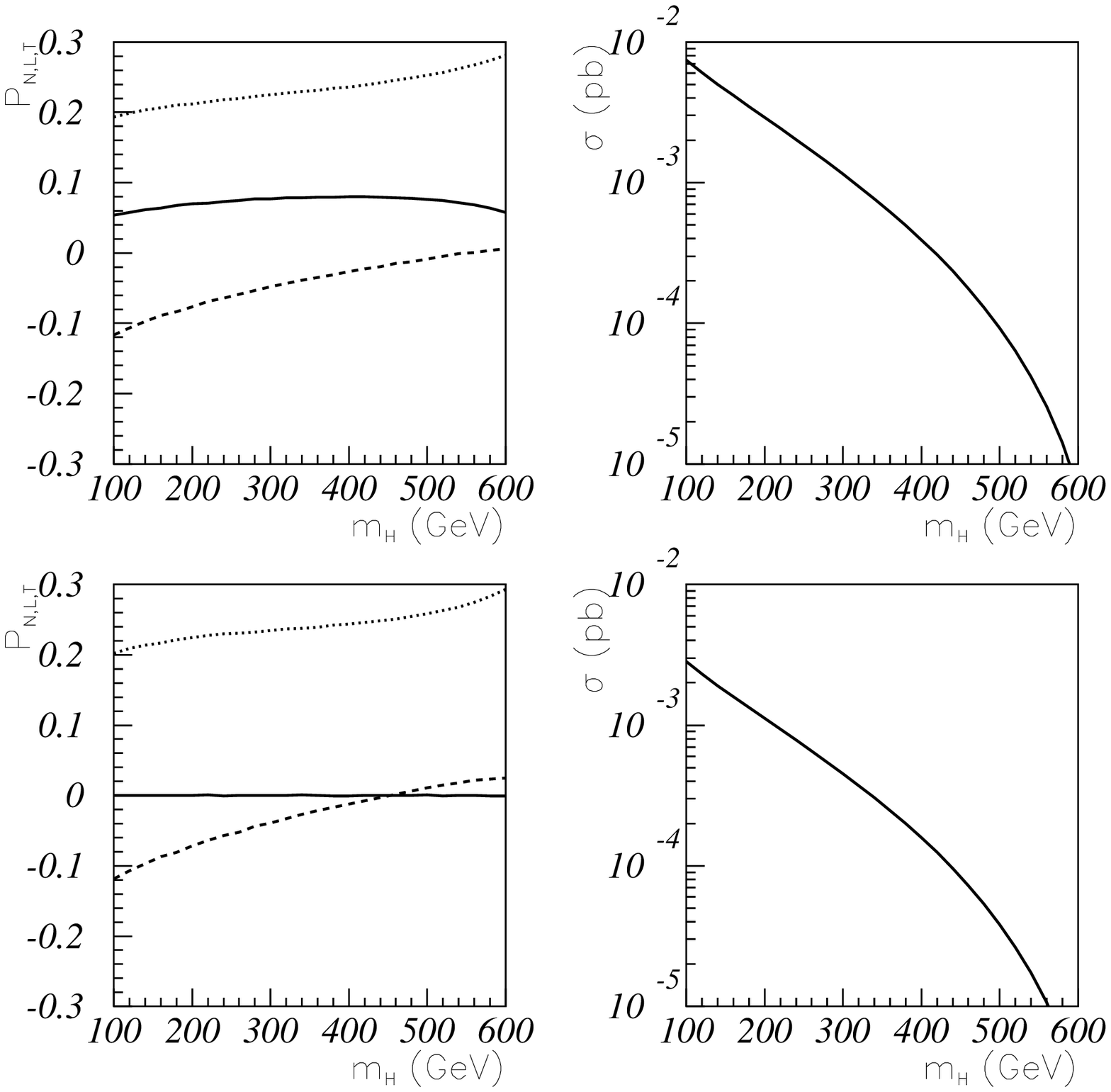}}
\caption{ 
Polarizations of top quark and the cross section as function
of Higgs mass for $e^+ e^- \rightarrow
t \bar t h$, where $\sqrt{s}=1$ TeV, the upper two figures
represent those in 2HDM and the lower in SM. 
The solid, dashed and dot-dashed lines represent the normal, logitudal
and transverse polarization of the top quark respectively, and for 2HDM, parameters
 are $\tan\beta=0.5$, $\alpha_1=\pi/4$,
$\alpha_2=\pi/4$, $\alpha_3=0$.
}
\label{fig1}
\end{figure}

\begin{figure}
\epsfxsize=12 cm
\centerline{\epsffile{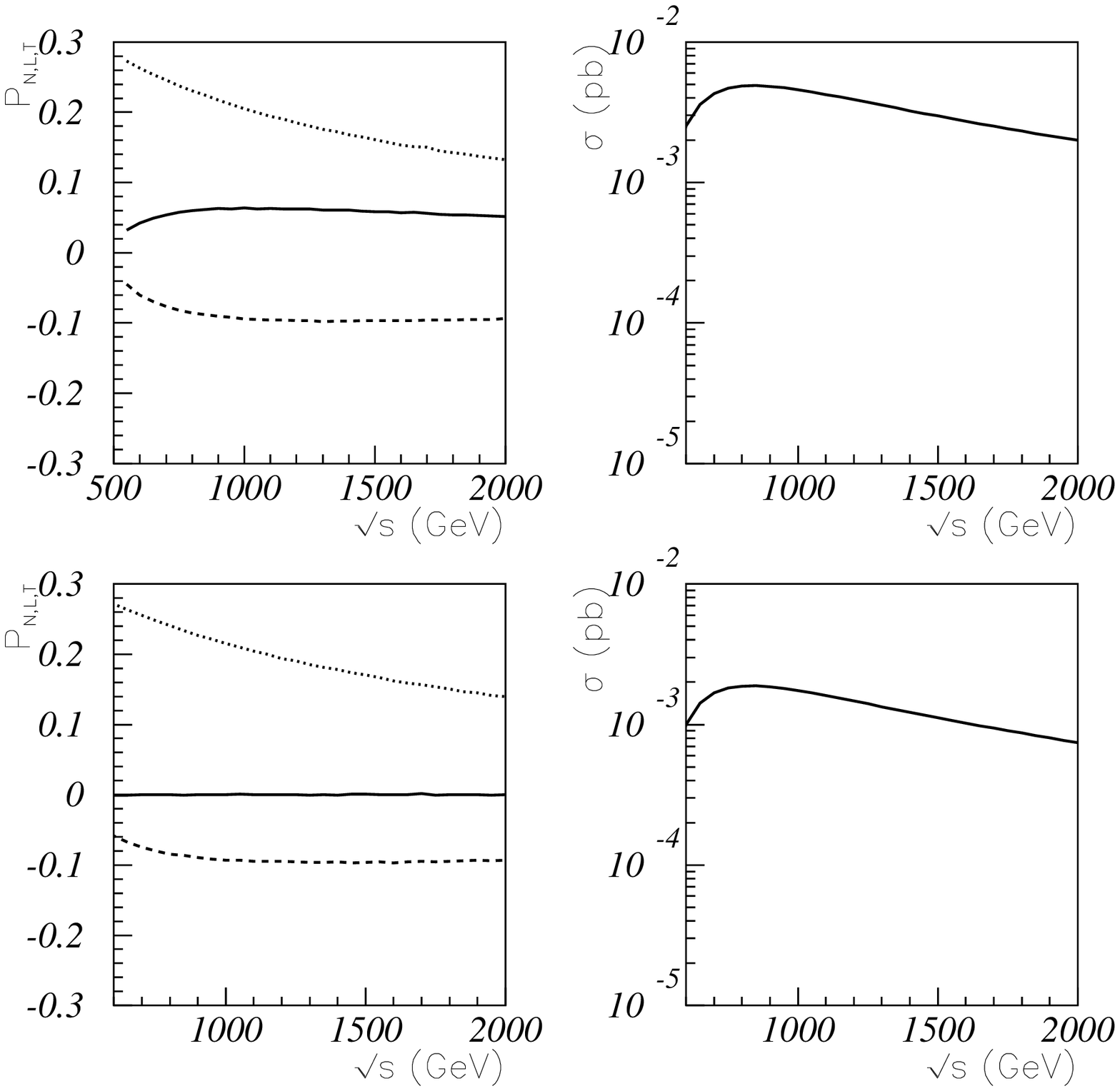}}
\caption{
Polarizations of top quark and the cross section as function
of center-of-mass-energy of $e^+ e^-$ for $e^+ e^- \rightarrow
t \bar t h$, where $m_h=150$ GeV. Other parameters and captions
are same as Fig. \ref{fig1}.}
\label{fig2}
\end{figure}

\begin{figure}
\epsfxsize=12 cm
\centerline{\epsffile{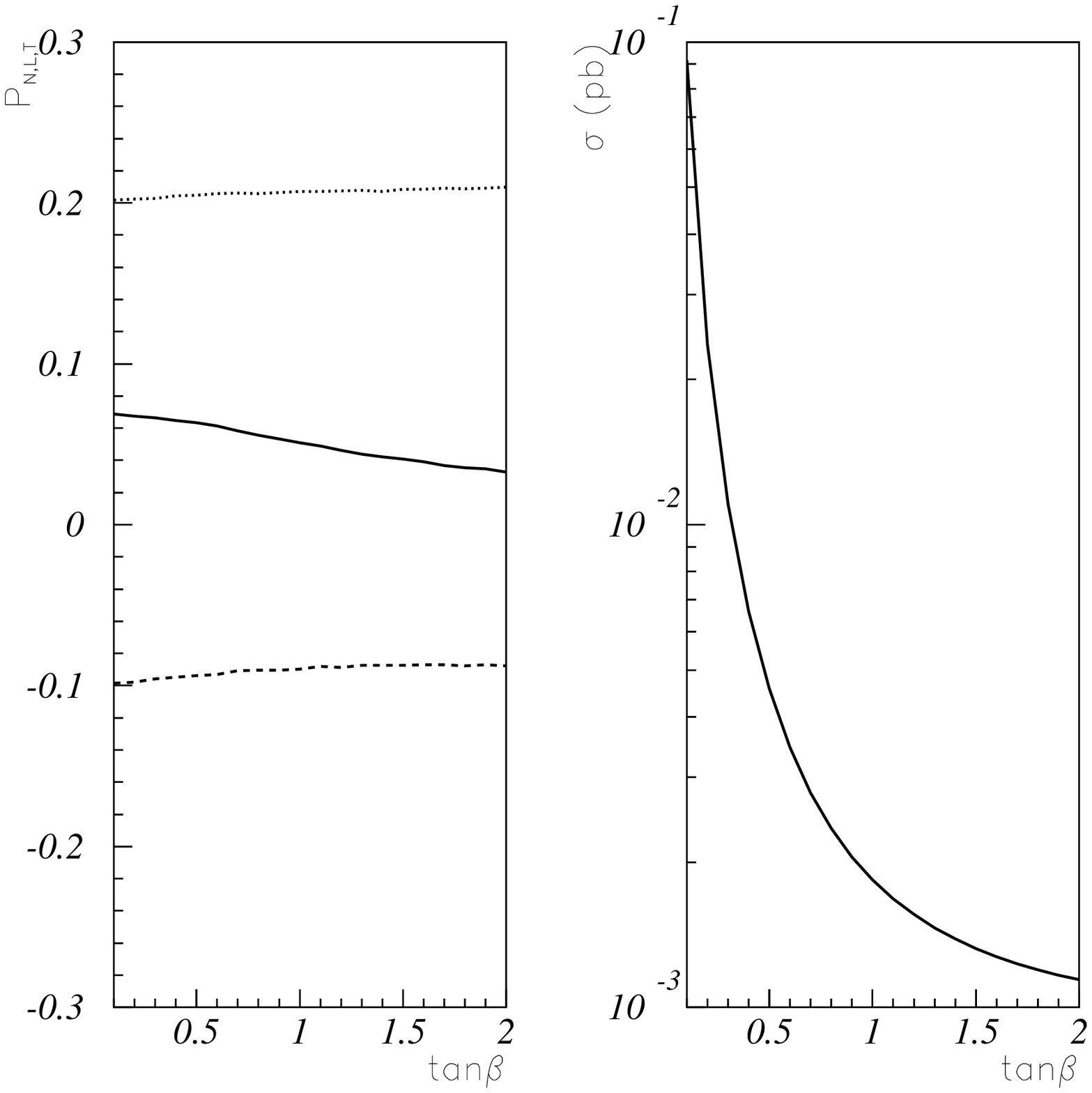}}
\caption{
Polarizations of top quark and  the cross section as function
of $\tan\beta$
for $e^+ e^- \rightarrow
t \bar t h$,
 where center-of-mass-energy of $e^+ e^-$ is 1 TeV and
$m_h=150$ GeV. Other model parameters 
are same as Fig. \ref{fig1}.}
\label{fig3}
\end{figure}

\begin{figure}
\epsfxsize=12 cm
\centerline{\epsffile{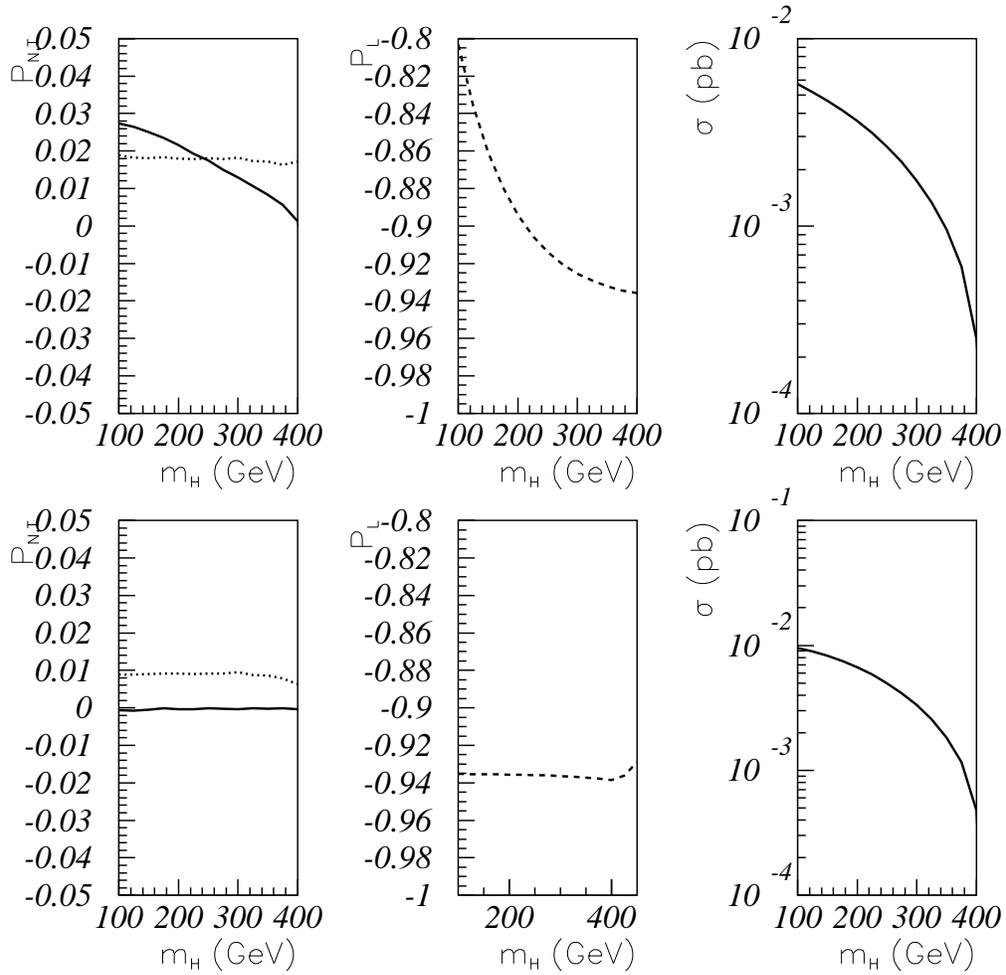}}
\caption{
Polarizations of bottom quark and the cross section as function
of Higgs mass for $e^+ e^- \rightarrow
b \bar b h$, where $\sqrt{s}=0.5$ TeV, the upper three figures
represent those in 2HDM and the lower in SM. 
For 2HDM, parameters are $\tan\beta=50$, $\alpha_1=\pi/4$,
$\alpha_2=\pi/4$, $\alpha_3=0$, and
for left column, 
solid and dashed lines represent the normal
and transverse polarization of the bottom quark respectively.}
\label{fig4}
\end{figure}

\begin{figure}
\epsfxsize=12 cm
\centerline{\epsffile{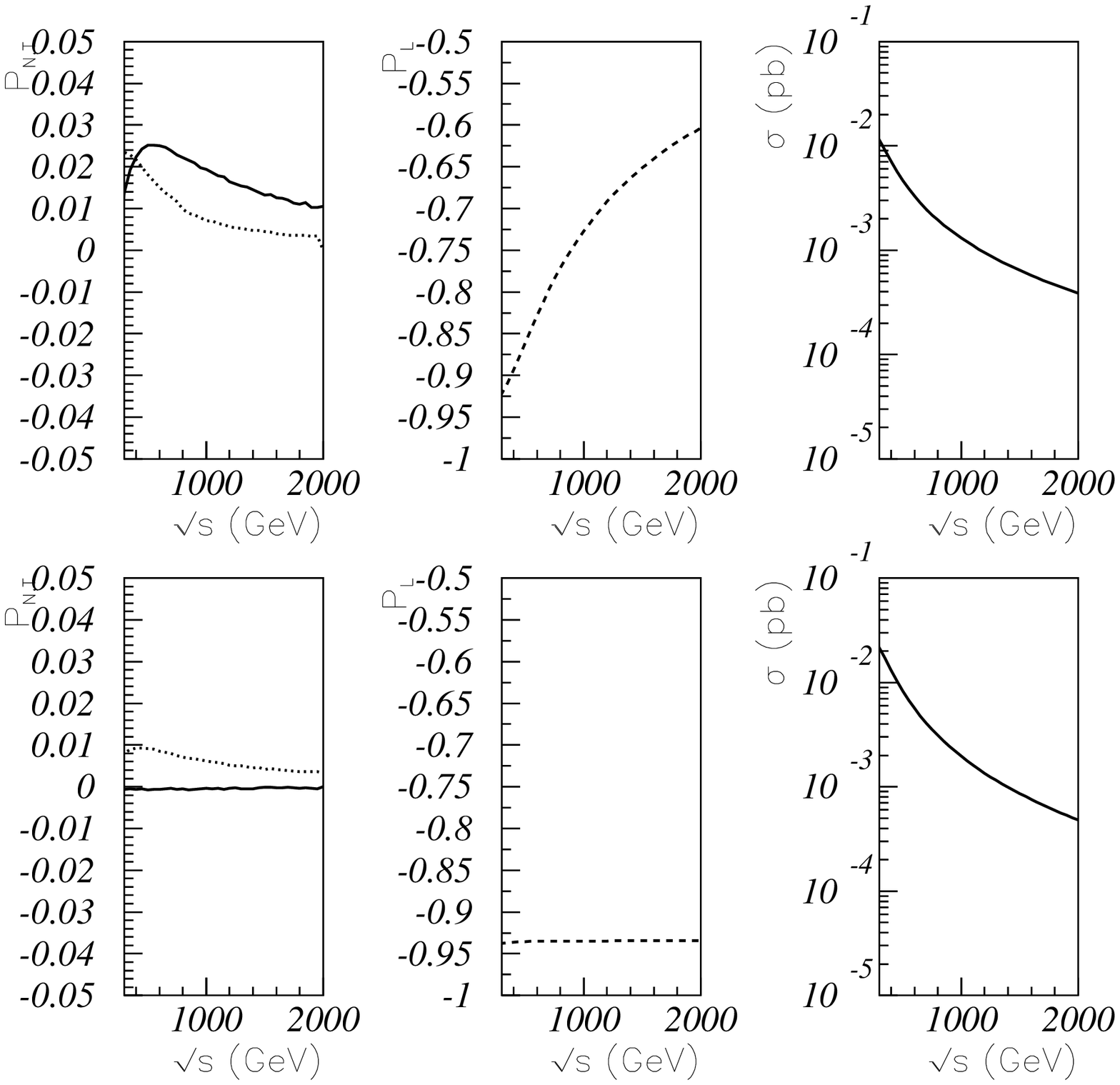}}
\caption{
Polarizations of bottom quark and the cross section as function
of center-of-mass-energy of $e^+ e^-$ for $e^+ e^- \rightarrow
b \bar b h$, where $m_h=150$ GeV. Other parameters and captions
are same as Fig. \ref{fig4}.}
\label{fig5}
\end{figure}

\begin{figure}
\epsfxsize=12 cm
\centerline{\epsffile{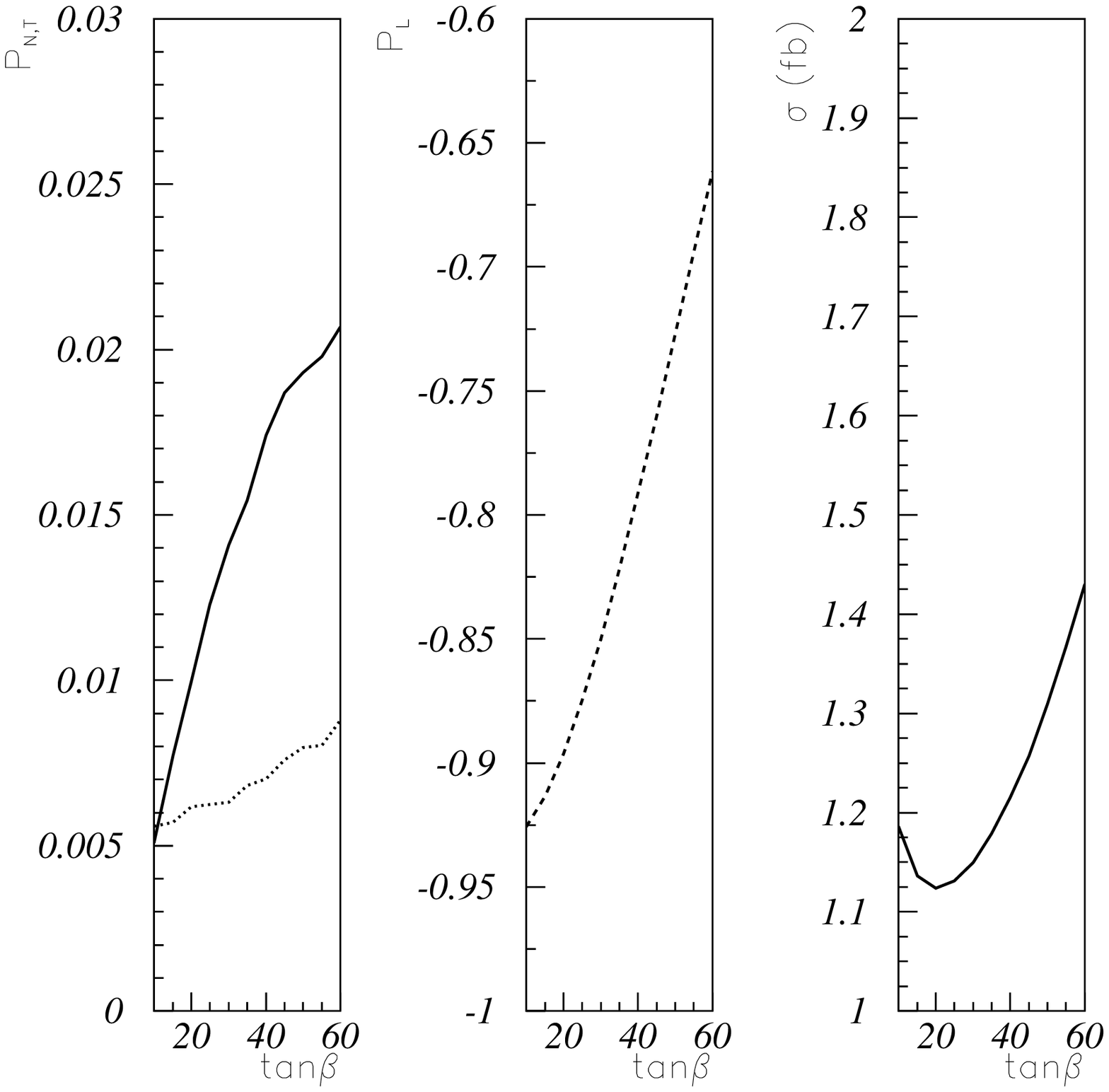}}
\caption{
Polarizations of bottom quark and the cross section as function
of $\tan\beta$
for $e^+ e^- \rightarrow
b \bar b h$,
 where center-of-mass-energy of $e^+ e^-$ is 1 TeV and
$m_h=150$ GeV. Other model parameters 
are same as Fig. \ref{fig4}.}
\label{fig6}
\end{figure}


\end{document}